%% file: phis-Bs2Jpsipipi-paper.tex
\pdfoutput=1
\documentclass[12pt,a4paper]{article}

\usepackage{lineno}  
\usepackage{graphicx}  
\usepackage{xspace} 
\usepackage{color}
\usepackage{colortbl}
\usepackage{amsmath} 
\usepackage{ifthen} 
\usepackage{multirow}

\newboolean{pdflatex}
\setboolean{pdflatex}{true} 
\newboolean{articletitles}
\setboolean{articletitles}{true} 

\newboolean{uprightparticles}
\setboolean{uprightparticles}{false} 

\usepackage{amssymb}
\usepackage{amsfonts}
\usepackage{upgreek} 

\usepackage{hyperref}    

\usepackage[all]{hypcap} 
\usepackage{afterpage}

\graphicspath{{fig/};{figures/}}

\def\lhcb {\mbox{LHCb}\xspace}
\def\ux85 {\mbox{UX85}\xspace}

\ifthenelse{\boolean{uprightparticles}}%
{

 \def\Ppsi        {\ensuremath{\uppsi}\xspace}

 \def\PDelta      {\ensuremath{\Delta}\xspace}
 \def\PXi      {\ensuremath{\Xi}\xspace}
 \def\PLambda      {\ensuremath{\Lambda}\xspace}
 \def\PSigma      {\ensuremath{\Sigma}\xspace}
 \def\POmega      {\ensuremath{\Omega}\xspace}
 \def\PUpsilon      {\ensuremath{\Upsilon}\xspace}

 \def\PB      {\ensuremath{\mathrm{B}}\xspace}
 
 \def\PD      {\ensuremath{\mathrm{D}}\xspace}

 \def\PJ      {\ensuremath{\mathrm{J}}\xspace}
 \def\PK      {\ensuremath{\mathrm{K}}\xspace}

 \def\Pb      {\ensuremath{\mathrm{b}}\xspace}
 \def\Pc      {\ensuremath{\mathrm{c}}\xspace}

 \def\Pi      {\ensuremath{\mathrm{i}}\xspace}

 \def\Ps      {\ensuremath{\mathrm{s}}\xspace}

}
{

 \def\Ppsi        {\ensuremath{\psi}\xspace}
 
 \mathchardef\PDelta="7101
 \mathchardef\PXi="7104
 \mathchardef\PLambda="7103
 \mathchardef\PSigma="7106
 \mathchardef\POmega="710A
 \mathchardef\PUpsilon="7107
 
 \def\PB      {\ensuremath{B}\xspace}
 
 \def\PD      {\ensuremath{D}\xspace}

 \def\PJ      {\ensuremath{J}\xspace}
 \def\PK      {\ensuremath{K}\xspace}

 \def\Pb      {\ensuremath{b}\xspace}
 \def\Pc      {\ensuremath{c}\xspace}

 \def\Pi      {\ensuremath{i}\xspace}

 \def\Ps      {\ensuremath{s}\xspace}

}

\makeatletter
\ifcase \@ptsize \relax
  \newcommand{\miniscule}{\@setfontsize\miniscule{4}{5}}
\or
  \newcommand{\miniscule}{\@setfontsize\miniscule{5}{6}}
\or
  \newcommand{\miniscule}{\@setfontsize\miniscule{5}{6}}
\fi
\makeatother

\def\squark    {\ensuremath{\Ps}\xspace}

\def\cquark    {\ensuremath{\Pc}\xspace}

\def\bquark    {\ensuremath{\Pb}\xspace}

\def\kaon  {\ensuremath{\PK}\xspace}
  \def\Kbar  {\kern 0.2em\overline{\kern -0.2em \PK}{}\xspace}

\def\Kz    {\ensuremath{\kaon^0}\xspace}
\def\Kzb   {\ensuremath{\Kbar^0}\xspace}
\def\KzKzb {\ensuremath{\Kz \kern -0.16em \Kzb}\xspace}
\def\Kp    {\ensuremath{\kaon^+}\xspace}
\def\Km    {\ensuremath{\kaon^-}\xspace}

\def\KpKm  {\ensuremath{\Kp \kern -0.16em \Km}\xspace}

\def\Kstarz  {\ensuremath{\kaon^{*0}}\xspace}
\def\Kstarzb {\ensuremath{\Kbar^{*0}}\xspace}

  \def\Dbar    {\kern 0.2em\overline{\kern -0.2em \PD}{}\xspace}
\def\D       {\ensuremath{\PD}\xspace}

\def\Dz      {\ensuremath{\D^0}\xspace}
\def\Dzb     {\ensuremath{\Dbar^0}\xspace}
\def\DzDzb   {\ensuremath{\Dz {\kern -0.16em \Dzb}}\xspace}
\def\Dp      {\ensuremath{\D^+}\xspace}
\def\Dm      {\ensuremath{\D^-}\xspace}

\def\DpDm    {\ensuremath{\Dp {\kern -0.16em \Dm}}\xspace}

\def\B       {\ensuremath{\PB}\xspace}
  \def\Bbar    {\kern 0.18em\overline{\kern -0.18em \PB}{}\xspace}

\def\Bz      {\ensuremath{\B^0}\xspace}
\def\Bzb     {\ensuremath{\Bbar^0}\xspace}

\def\Bs      {\ensuremath{\B^0_\squark}\xspace}
\def\Bsb     {\ensuremath{\Bbar^0_\squark}\xspace}

\def\jpsi     {\ensuremath{{\PJ\mskip -3mu/\mskip -2mu\Ppsi\mskip 2mu}}\xspace}

  \def\Y#1S{\ensuremath{\PUpsilon{(#1S)}}\xspace}

\def\L {\ensuremath{\PLambda}\xspace}
\def\Lbar {\ensuremath{\kern 0.1em\overline{\kern -0.1em\Lambda\kern -0.05em}\kern 0.05em{}}\xspace}

\def\Lb      {\ensuremath{\L^0_\bquark}\xspace}
\def\Lbbar   {\ensuremath\overline{{\Lb}}\xspace}

\def\to                 {\ensuremath{\rightarrow}\xspace}

\def\CP                {\ensuremath{C\!P}\xspace}

\newcommand{\dms}{\ensuremath{\Delta m_{\squark}}\xspace}

\newcommand{\DGs}{\ensuremath{\Delta\Gamma_{\squark}}\xspace}

\newcommand{\Gs}{\ensuremath{\Gamma_{\squark}}\xspace}

\def\AT#1     {\ensuremath{A_{\mathrm{T}}^{#1}}\xspace}           

\def\C#1      {\ensuremath{\mathcal{C}_{#1}}\xspace}                       
\def\Cp#1     {\ensuremath{\mathcal{C}_{#1}^{'}}\xspace}                    
\def\Ceff#1   {\ensuremath{\mathcal{C}_{#1}^{\mathrm{(eff)}}}\xspace}        
\def\Cpeff#1  {\ensuremath{\mathcal{C}_{#1}^{'\mathrm{(eff)}}}\xspace}       
\def\Ope#1    {\ensuremath{\mathcal{O}_{#1}}\xspace}                       
\def\Opep#1   {\ensuremath{\mathcal{O}_{#1}^{'}}\xspace}                    

\newcommand{\tev}{\ensuremath{\mathrm{\,Te\kern -0.1em V}}\xspace}
\newcommand{\gev}{\ensuremath{\mathrm{\,Ge\kern -0.1em V}}\xspace}
\newcommand{\mev}{\ensuremath{\mathrm{\,Me\kern -0.1em V}}\xspace}
\newcommand{\kev}{\ensuremath{\mathrm{\,ke\kern -0.1em V}}\xspace}
\newcommand{\ev}{\ensuremath{\mathrm{\,e\kern -0.1em V}}\xspace}
\newcommand{\gevc}{\ensuremath{{\mathrm{\,Ge\kern -0.1em V\!/}c}}\xspace}
\newcommand{\mevc}{\ensuremath{{\mathrm{\,Me\kern -0.1em V\!/}c}}\xspace}
\newcommand{\gevcc}{\ensuremath{{\mathrm{\,Ge\kern -0.1em V\!/}c^2}}\xspace}
\newcommand{\gevgevcccc}{\ensuremath{{\mathrm{\,Ge\kern -0.1em V^2\!/}c^4}}\xspace}
\newcommand{\mevcc}{\ensuremath{{\mathrm{\,Me\kern -0.1em V\!/}c^2}}\xspace}

\def\m    {\ensuremath{\rm \,m}\xspace}

\def\mum  {\ensuremath{\,\upmu\rm m}\xspace}

\def\invfb   {\ensuremath{\mbox{\,fb}^{-1}}\xspace}

\def\ps   {\ensuremath{{\rm \,ps}}\xspace}
\def\fs   {\ensuremath{\rm \,fs}\xspace}

\def\gsim{{~\raise.15em\hbox{$>$}\kern-.85em
          \lower.35em\hbox{$\sim$}~}\xspace}
\def\lsim{{~\raise.15em\hbox{$<$}\kern-.85em
          \lower.35em\hbox{$\sim$}~}\xspace}

\newcommand{\Real}{\ensuremath{\mathcal{R}e}\xspace}
\newcommand{\Imag}{\ensuremath{\mathcal{I}m}\xspace}

\def\pt         {\mbox{$p_{\rm T}$}\xspace}

\def\photos     {\mbox{\textsc{Photos}}\xspace}
\def\evtgen     {\mbox{\textsc{EvtGen}}\xspace}
\def\pythia     {\mbox{\textsc{Pythia}}\xspace}

\def\geant      {\mbox{\textsc{Geant4}}\xspace}

\def\gauss      {\mbox{\textsc{Gauss}}\xspace}

\def\tell1  {TELL1\xspace}
\def\ukl1   {UKL1\xspace}

\def \A {{\cal A}}
\def \cAb {\overline{{\cal A}}}

\def \m {m_{hh}}
\def \angmu {\theta_{\jpsi}}
\def \angpi {\theta_{hh}}

\def \Bq {B_s^{0}}

\def \ch {\cosh \frac{\DGs t}{2}}
\def \sh {\sinh \frac{\DGs t}{2}}
\def \Ab {\overline{A}}
\def \cs {\cos(\dms t)}
\def \sn {\sin(\dms t)}

\DeclareRobustCommand{\optbar}[1]{\shortstack{{\miniscule (\rule[.5ex]{1.25em}{.18mm})}
  \\ [-.7ex] $#1$}}

\usepackage{epsfig,accents}
\newcommand*{\fancybar}{\scalebox{.4}{(}\raisebox{-1.7pt}{--}\scalebox{.4}{)}}
\newcommand*{\brabar}[1]{\accentset{\fancybar}{#1}}
\newcommand\BorBbar{\kern 0.18em\optbar{\kern -0.18em B}{}\xspace}
\newcommand\KorKbar{\kern 0.18em\optbar{\kern -0.18em K}{}\xspace}

\usepackage{cite}
\usepackage{mciteplus}
\begin{document}
\renewcommand{\thefootnote}{\fnsymbol{footnote}}
\setcounter{footnote}{1}
\begin{titlepage}

\belowpdfbookmark{Title page}{title}

\pagenumbering{roman}
\vspace*{-1.5cm}
\centerline{\large EUROPEAN ORGANIZATION FOR NUCLEAR RESEARCH (CERN)}
\vspace*{1.5cm}
\hspace*{-5mm}\begin{tabular*}{16cm}{lc@{\extracolsep{\fill}}r}
\vspace*{-12mm}\mbox{\!\!\!\includegraphics[width=.12\textwidth]{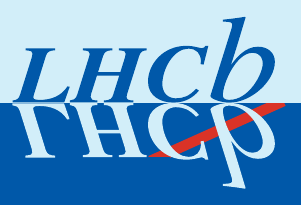}}& & \\
 & & CERN-PH-EP-2014-089\\
 & & LHCb-PAPER-2014-019\\  
 & & \today \\ 
 & & \\
\end{tabular*}

\vspace*{2.0cm}

{\bf\boldmath\Large
\begin{center}
Measurement of the \CP-violating phase $\phi_s$ in  $\Bsb\rightarrow J/\psi \pi^+\pi^-$ decays
\end{center}
}

\vspace*{1.0cm}
\begin{center}
\normalsize {
The LHCb collaboration\footnote{Authors are listed on the following pages.}
}
\end{center}

\begin{abstract}
  \noindent
The mixing-induced \CP-violating phase $\phi_s$ in \Bs and $\Bsb$ decays is measured using the $\jpsi \pi^+\pi^-$ final state in data, taken from  3\,fb$^{-1}$ of integrated luminosity, collected with the LHCb detector in 7 and 8~TeV centre-of-mass $pp$ collisions at the LHC.  A time-dependent  flavour-tagged amplitude analysis, allowing for direct \CP violation, yields a value for the phase $\phi_s=70\pm 68\pm 8$\,mrad. This result is consistent with the Standard Model expectation and previous measurements.
\end{abstract}

\vspace*{2.0cm}
\vspace{\fill}


\vspace*{1.0cm}
\begin{center}
\hspace*{6mm}Submitted to Physics Letter B\\
\end{center}
\vspace{\fill}

{\footnotesize
\centerline{\copyright~CERN on behalf of the \lhcb collaboration, license \href{http://creativecommons.org/licenses/by/3.0/}{CC-BY-3.0}.}}
\vspace*{2mm}



\end{titlepage}

\newpage
\setcounter{page}{2}
\mbox{~}

\input{LHCb_HD_authorlist_2014-04-08.tex}

\cleardoublepage

\renewcommand{\thefootnote}{\arabic{footnote}}
\setcounter{footnote}{0}



\pagestyle{plain} 
\setcounter{page}{1}
\pagenumbering{arabic}



%
\clearpage

\renewcommand{\thefootnote}{\arabic{footnote}}
\setcounter{footnote}{0}










\section{Introduction}
\label{sec:Introduction}
One of the most sensitive ways of detecting the presence of heretofore unseen particles or forces is through the observation of effects they may have on \CP-violating decays of neutral $B$ mesons \cite{Dunietz:2000cr}. Measurements of \CP violation through the interference of \Bs mixing and decay amplitudes are particularly sensitive because the Standard Model (SM) prediction of the \CP-violating phase is very small and accurate in quark level $b\to c\overline{c}s$ transitions, with $\phi_s^{\rm SM}\equiv -2{\rm arg}\left(-\frac{V_{ts}V_{tb}^*}{V_{cs}V_{cb}^*}\right)\!=\!-36.3_{-1.5}^{+1.6}$\,mrad, ignoring subleading corrections from Penguin amplitudes~\cite{Charles:2011va}. Initial measurements of $\phi_s$ at the Tevatron indicated possible large values inconsistent with the SM expectation \cite{Aaltonen:2007he,*Abazov:2008af,*CDF:2011af,*Abazov:2011ry}, while LHCb measurements using both $\BorBbar^0_{\!\!s}\to\jpsi\phi$ and $\BorBbar^0_{\!\!s}\to\jpsi\pi^+\pi^-$ decays from 1\invfb of integrated luminosity were consistent with the SM value \cite{Aaij:2013oba,LHCb:2012ad}, as were more recent results from CDF~\cite{Aaltonen:2012ie}, and ATLAS~\cite{Aad:2012kba}.

In this Letter, we present a new measurement of $\phi_s$ in $\BorBbar^0_{\!\!s}\to\jpsi\pi^+\pi^-$ decays using data taken from an
integrated luminosity of 3\invfb, obtained from $pp$ collisions at the LHC. One-third of the data was collected at a centre-of-mass energy of 7\tev, and the remainder at 8\tev. In the previous study we used the result of our amplitude analysis~\cite{LHCb:2012ae}, which showed that the \CP-odd  component of the decay was larger than $97.7$\% at 95\% confidence level (CL). Here we perform a more sophisticated amplitude analysis \cite{Zhang:2012zk}, which uses an additional angular variable, and thereby directly determines
the \CP-odd and \CP-even components. Previously it
was found that  five interfering $\pi^+\pi^-$ states are required to describe the decay: $f_0(980), f_0(1500), f_0(1790), f_2(1270)$, and $f_2^{\prime}(1525)$ \cite{Aaij:2014emv}. An alternative model including these states and a nonresonant $\jpsi \pi^+\pi^-$ component also provides a good description of the data; the limit on the  \CP-even component is unchanged. The $\jpsi f_0(980)$ final state was suggested as being a useful final state for measuring $\phi_s$ as it is a \CP-eigenstate ~\cite{Stone:2008ak} and inspired these studies. Subsequently, it was suggested that the $f_0(980)$ resonance might be formed of tetraquarks \cite{Fleischer:2011au}, and could then provide an additional SM contribution to $\phi_s$ beyond that originally expected. Studies of $\Bzb\to\jpsi\pi^+\pi^-$ decays \cite{Aaij:2014zda}, indicate that the light scalar mesons are familiar $q\overline{q}$ states \cite{Stone:2013eaa}, so this concern has been ameliorated.

The method used here allows the measurement of the \CP-violating phase $\phi_s$, without any assumption on the \CP content, by measuring simultaneously the \CP-even and \CP-odd decay amplitudes and $\phi_s$.

\section{\boldmath Decay rates for $\Bsb \to \jpsi h^+ h^-$}

The differential decay rates for $\BorBbar^0_{\!\!s}\to\jpsi h^+h^-$, allowing for possible direct \CP violation, can be written in terms of the decay time $t$, and the decay amplitudes $\A$ and $\cAb$ as \cite{Nierste:2009wg,*Bigi:2000yz}
\begin{eqnarray}\label{Eq-t}
\Gamma(t) =\quad\quad
  {\cal N} e^{-\Gs t}\left\{\frac{|\A|^2+|\cAb|^2}{2}\ch  + \frac{|\A|^2-|\cAb|^2}{2}\cs\right.\quad\quad\nonumber\\
- \left.\Real(\A^*\cAb)\sh  -  \Imag(\A^*\cAb)\sn\right\},
\end{eqnarray}
\begin{eqnarray}
\overline{\Gamma}(t) =
 \left|\frac{p}{q}\right|^2{\cal N}  e^{-\Gs t}\left\{\frac{|\A|^2+|\cAb|^2}{2}\ch  - \frac{|\A|^2-|\cAb|^2}{2}\cs\right.\quad\quad\nonumber\\
- \left.\Real(\A^*\cAb)\sh  +  \Imag(\A^*\cAb)\sn\right\},\label{Eqbar-t}
\end{eqnarray}
where $\DGs \equiv \Gamma_L-\Gamma_H$ is the decay width difference between the light and the heavy mass eigenstates, $\dms \equiv m_H-m_L$ is the mass difference, $\Gs \equiv (\Gamma_L+\Gamma_H)/2$ is the average width, and $\cal N$ is a constant.  The complex parameters $q$ and $p$ are used to relate the mixing between the mass and flavour eigenstates. The decay amplitudes are defined as ${\cal A}\equiv A_f$ and $\cAb\equiv \frac{q}{p} \overline{A}_f$, where $A_f$ ($\overline{A}_f$) is the total amplitude of $\Bs$ ($\Bsb$) $\to \jpsi h^+h^-$ decays at time $t=0$.

The total amplitude $A_f$ ($\overline{A}_f$) is taken to be the sum over individual $\pi^+\pi^-$ resonant transversity amplitudes \cite{Dighe:1995pd}, and possibly one nonresonant amplitude, labelled as ${A}_i$ ($\overline{A}_i$). By introducing the parameter $\lambda_i \equiv \frac{q}{p}\frac{\Ab_i}{A_i}$, relating \CP violation in the interference between mixing and decay associated with the state $i$, the amplitudes ${\cal A}$ and $\cAb$ can be further expressed as ${\cal A}=\sum A_i$ and $\cAb =\sum \lambda_i A_i$, the sums of the individual $\BorBbar^0_{\!\!s}$ amplitudes.

For $\jpsi$ decays to $\mu^+\mu^-$ final states, these amplitudes are themselves functions of four variables: the $\pi^+\pi^-$ invariant mass $\m = m(\pi^+\pi^-)$,
and the three angles $\Omega$, defined in the helicity basis. These consist of
the angle between the $\mu^+$ direction in the $\jpsi$ rest frame with respect to the $\jpsi$ direction in the $\BorBbar^0_{\!\!s}$ rest frame $\angmu$,  the angle between the $h^+$ direction in the $h^+h^-$ rest frame with respect to the $h^+h^-$ direction in the $\Bq$ rest frame $\angpi$,
and the angle between the $\jpsi$ and $h^+h^-$ decay planes in the $\Bsb$ rest frame $\chi$ \cite{Zhang:2012zk,Aaij:2013oba}.

Assuming that any possible \CP violation in the decay is the same for all amplitudes, $\lambda\equiv \eta_i\lambda_i$ is common for all amplitudes, where $\eta_i$ is the \CP eigenvalue of the transversity state $i$. The \CP-violating phase $\phi_s$ is defined by $\phi_s\equiv -\arg(\lambda)$~\cite{Aaij:2013oba}, and appears in the term containing $\A^*\cAb$. The explicit forms of $|\brabar{\cal A}(\m,\Omega)|^2$ and $\A^*(\m,\Omega)\cAb(\m,\Omega)$ in Eqs.~(\ref{Eq-t}) and (\ref{Eqbar-t}) as functions of $\m$ and $\Omega$ are given in Ref.~\cite{Zhang:2012zk}.

The factor $|p/q|^2$ is related to the flavour-specific \CP-violating asymmetry $a_{\rm sl}^s$ as
\begin{equation}
a_{\rm sl}^s \equiv \frac{|p/q|^2-|q/p|^2}{|p/q|^2+|q/p|^2}\approx |p/q|^2-1.
\end{equation}
LHCb measured  $a_{\rm sl}^s=(-0.06\pm0.50\pm0.36)\%$~\cite{Aaij:2013gta}, corresponding to $|p/q|^2=0.9994\pm0.0062$.  Thus, we take $|p/q|^2=1$ for what follows.

\section{The LHCb detector and event selection}
The \lhcb detector~\cite{LHCb-det} is a single-arm forward
spectrometer covering the \mbox{pseudorapidity} range $2<\eta <5$,
designed for the study of particles containing \bquark or \cquark
quarks. The detector includes a high-precision tracking system
consisting of a silicon-strip vertex detector surrounding the $pp$
interaction region, a large-area silicon-strip detector located
upstream of a dipole magnet with a bending power of about
$4{\rm\,Tm}$, and three stations of silicon-strip detectors and straw
drift tubes placed downstream.
The combined tracking system provides a momentum measurement\footnote{We use units where c=1.}  with
relative uncertainty that varies from 0.4\% at 5\,GeV to 0.6\% at 100\,GeV,
and impact parameter resolution of 20\mum for
tracks with large transverse momentum (\pt). Different types of charged hadrons are distinguished by information
from two ring-imaging Cherenkov detectors. 
Photon, electron and hadron candidates are identified by a calorimeter system consisting of scintillating-pad and preshower detectors, an electromagnetic
calorimeter and a hadronic calorimeter.
The trigger 
consists of a hardware stage, based on information from the calorimeter and muon systems, followed by a software stage, which applies a full event
reconstruction. Events selected for this analysis are triggered by a $\jpsi\to\mu^+\mu^-$ decay, where the $\jpsi$ is required at the software level to be consistent with coming from the decay of a $b$ hadron by use of either impact parameter requirements on the muons or detachment of the reconstructed $\jpsi$ decay position from the associated primary vertex. 

\begin{figure}[!t]
\begin{center}
\includegraphics[width=0.72\textwidth]{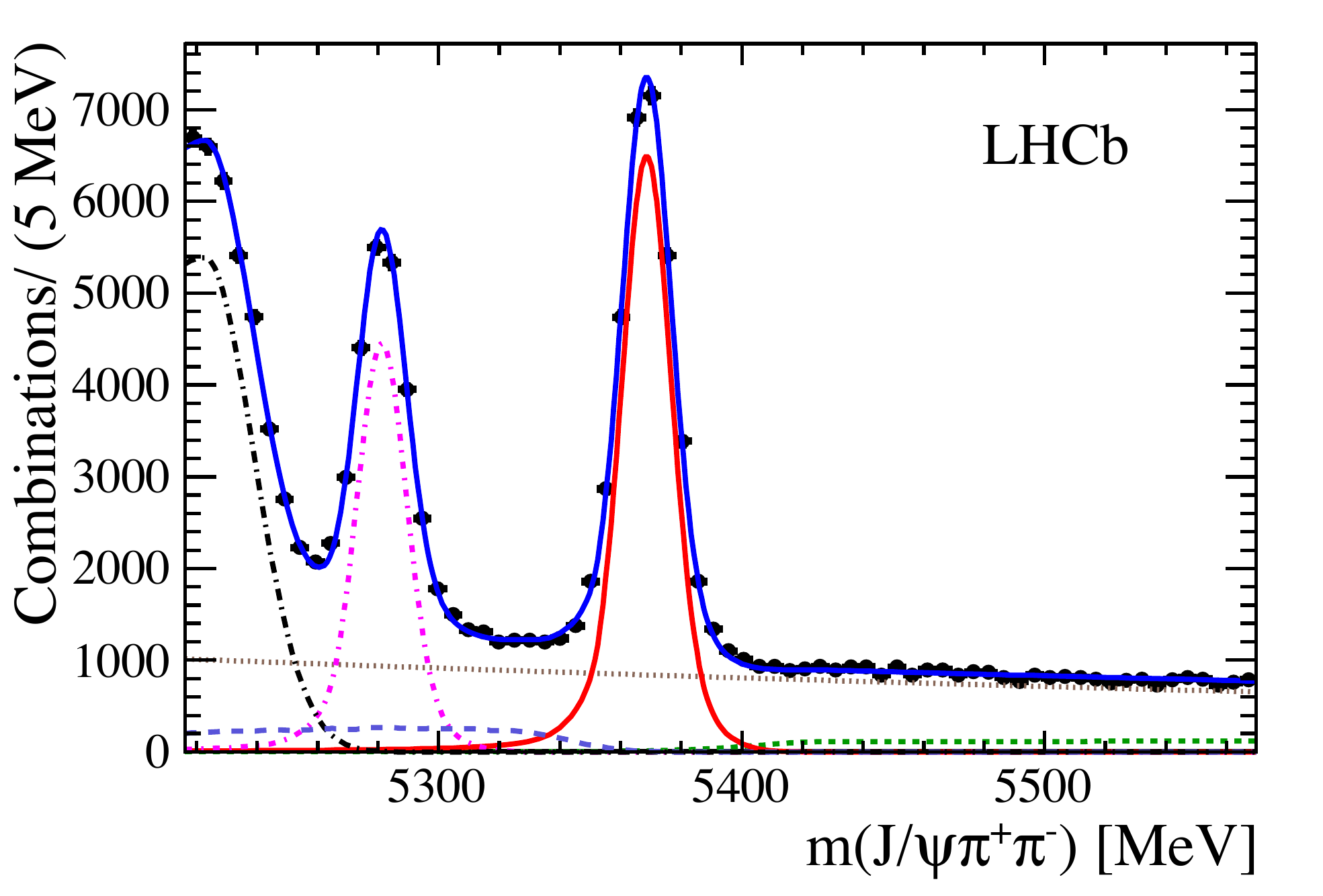}
\end{center}\label{fitmass}
\vskip -1cm
\caption{\small Invariant mass of $J/\psi \pi^+\pi^-$ combinations. The data are fitted with double Crystal Ball signal functions and several background functions. The (red) solid line shows the $\BorBbar^0_{\!\!s}$ signal, the (brown) dotted line shows the exponential combinatorial background, the (green) short-dashed line shows the $B^{\mp}$ background, the (magenta) dot-dashed  line shows the  $\BorBbar^0$ signal, the (light blue) dashed line is the sum of $\BorBbar^0_{\!\!s}\rightarrow J/\psi\eta'$, $\BorBbar^0_{\!\!s}\rightarrow J/\psi\phi$, $\phi\to\pi^+\pi^-\pi^0$ backgrounds, and the $\Lb\to \jpsi K^- p$ plus $\Lbbar\to\jpsi K^+\overline{p}$  reflections, the (black) dot-dashed line is the $\BorBbar^0\rightarrow J/\psi K^{\mp} \pi^{\pm}$ reflection and the (blue) solid line is the total.
}
\end{figure}

\begin{figure}[!t]
  \begin{center}
     \includegraphics[width=0.5\textwidth]{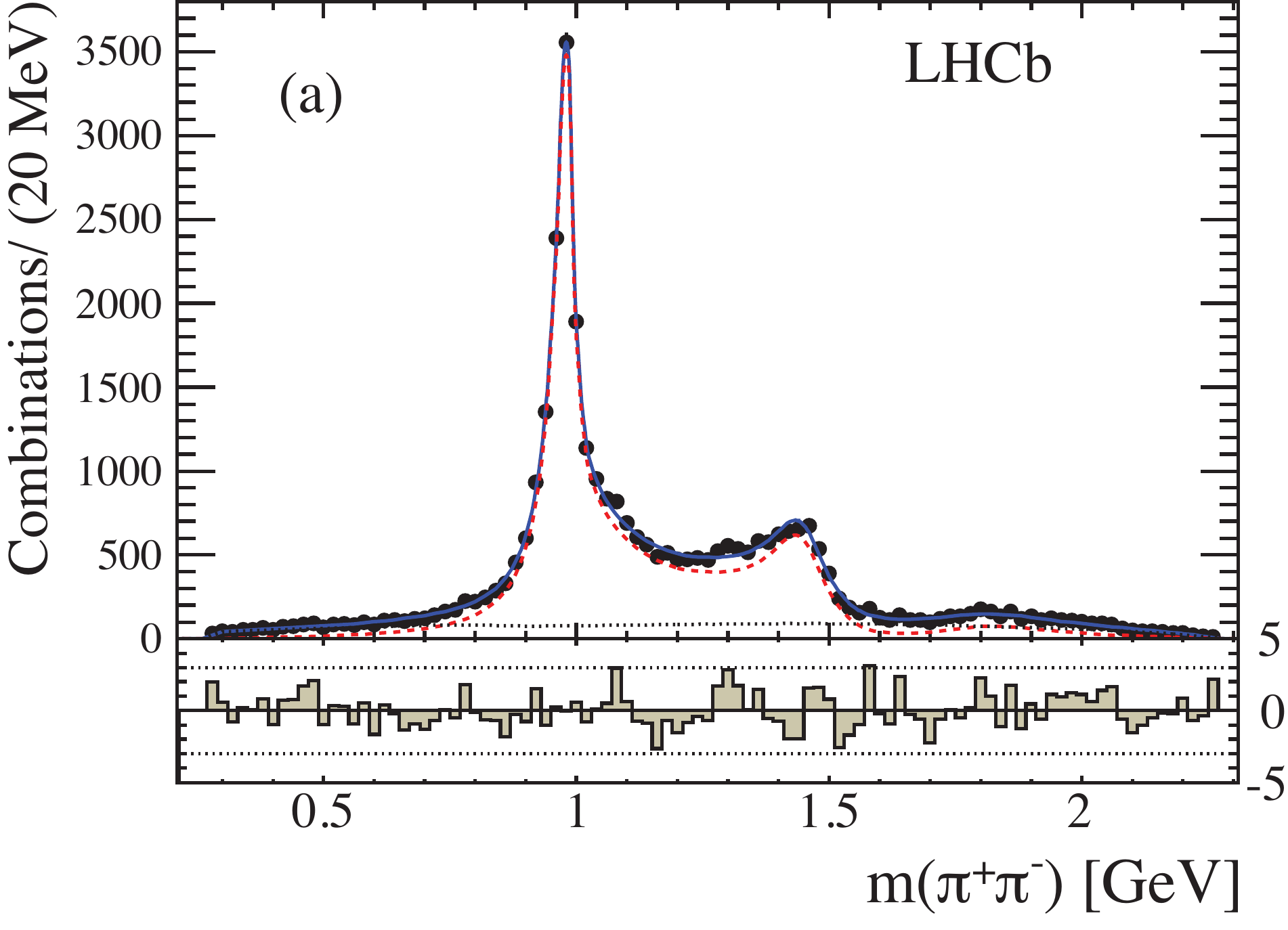}%
     \includegraphics[width=0.5\textwidth]{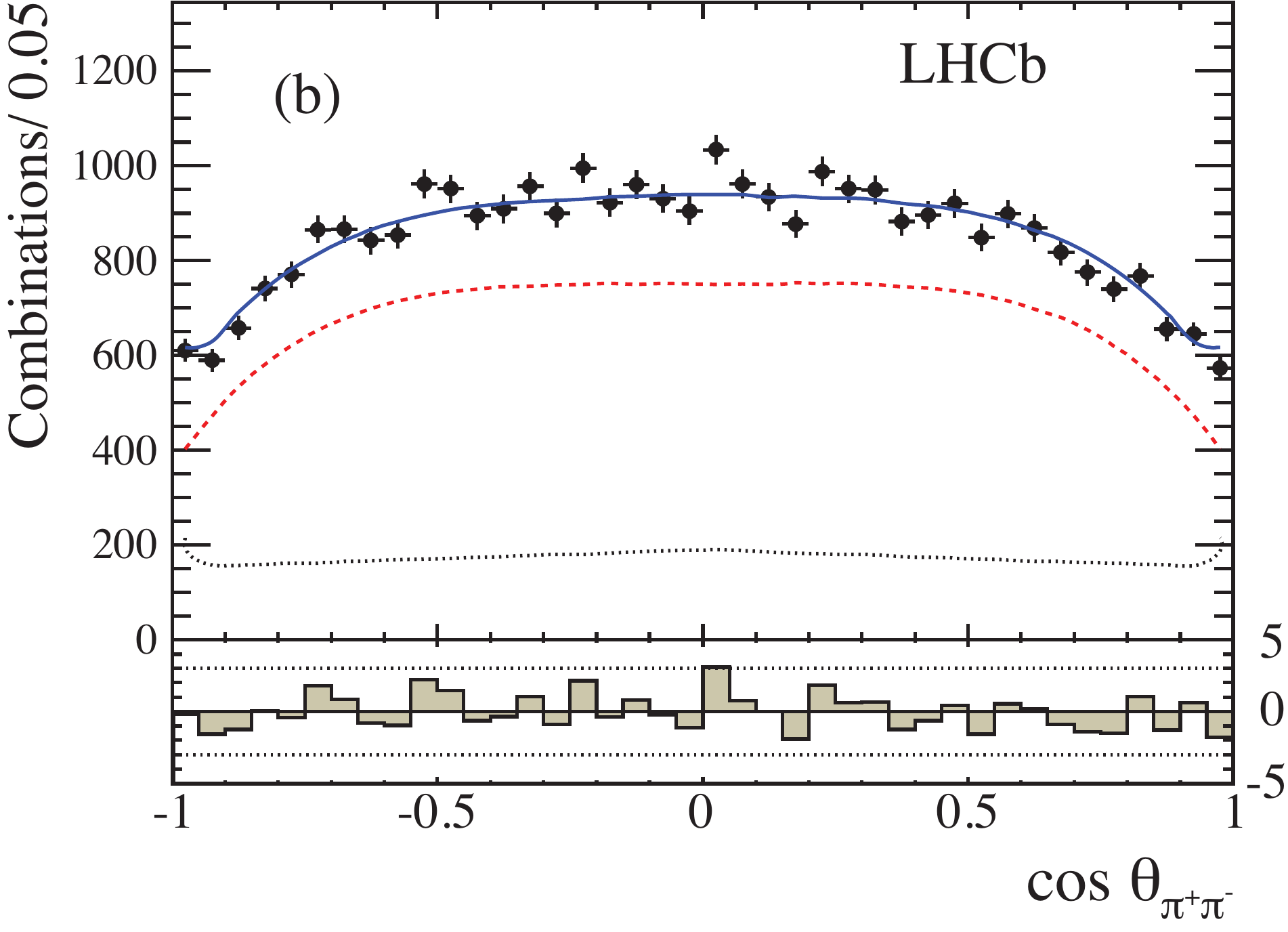}\\\vspace*{2mm}
      \includegraphics[width=0.5\textwidth]{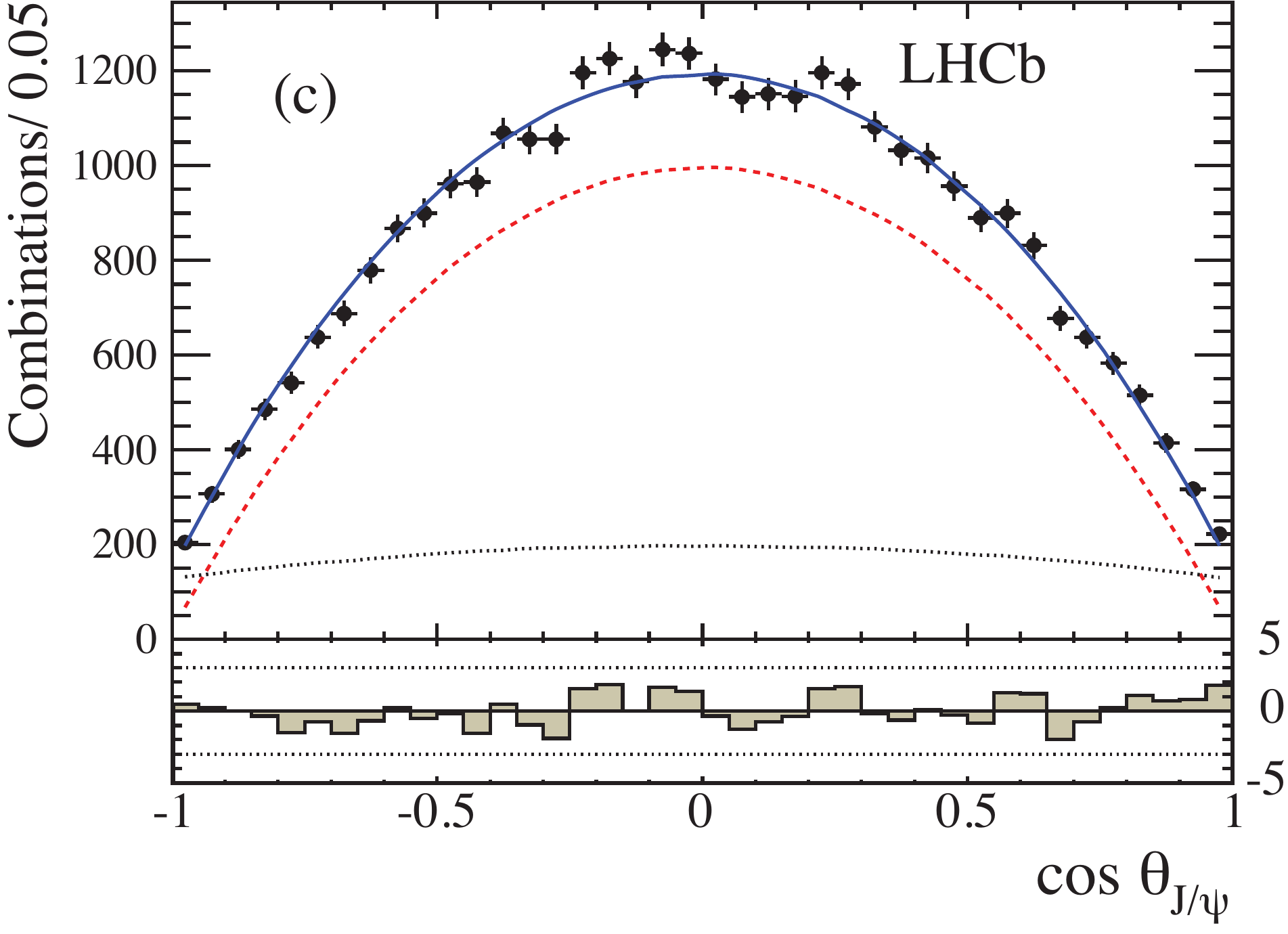}%
     \includegraphics[width=0.5\textwidth]{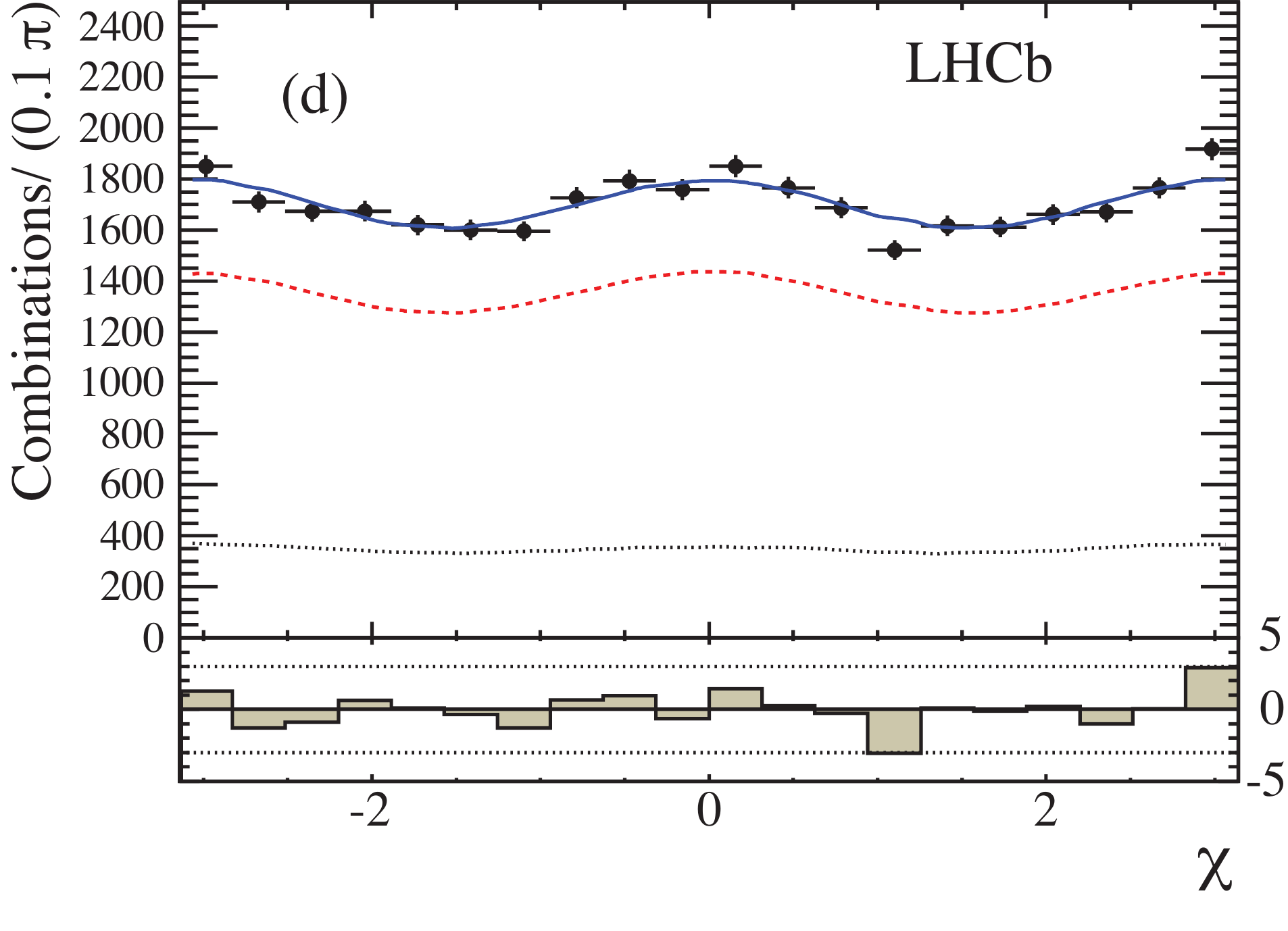}\\
    \caption{\small Projections of (a) $m(\pi^+\pi^-)$, (b) $\cos\theta_{\pi\pi}$, (c) $\cos \theta_{J/\psi}$ and (d) $\chi$~\cite{Aaij:2014emv}. The points with error bars are data, the signal fits are shown with (red) dashed lines, the background with a (black) dotted lines, and the (blue) solid lines represent the total fits. The difference between the data and the fits divided by the uncertainty on the data is shown below.}  \label{RM8}
  \end{center}
\end{figure}

A $\BorBbar^0_{\!\!s} \to \jpsi \pi^+\pi^-$ candidate is reconstructed by combining a $\jpsi \to \mu^+\mu^-$ candidate with two pions of opposite charge. The like-sign combinations $\jpsi\pi^\pm\pi^\pm$ are also reconstructed for background studies. Events are selected using a multivariate method that optimizes the ratio of signal squared to background events. The event selection is described in detail in the time-integrated amplitude analysis~\cite{Aaij:2014emv}.  The invariant mass distribution of $\jpsi\pi^+\pi^-$ combinations satisfying the event selection is shown in Fig.~\ref{fitmass}. Only the candidates within $\pm20$\mev of the $\Bsb$ mass peak are retained for the $\phi_s$ measurement; there are $27\,100\pm200$ signal events with a purity of 79.6\%. The integrated distributions of the four variables discussed above are shown in Fig.~\ref{RM8}.


Samples of simulated events are used to characterize signal and backgrounds.
In the simulation, $pp$ collisions are generated using
\pythia~\cite{Sjostrand:2006za,*Sjostrand:2007gs}
with a specific \lhcb
configuration~\cite{LHCb-PROC-2010-056}.  Decays of hadronic particles
are described by \evtgen~\cite{Lange:2001uf}, in which final state
radiation is generated using \photos~\cite{Golonka:2005pn}. The
interaction of the generated particles with the detector and its
response are implemented using the \geant
toolkit~\cite{Allison:2006ve, *Agostinelli:2002hh} as described in Ref.~\cite{LHCb-PROC-2011-006}.

\section{Likelihood construction}
We perform an unbinned maximum likelihood fit to the $\jpsi\pi^+\pi^-$ invariant mass $m$, the decay time $t$,  $m_{hh}$, and the three helicity angles $\Omega$, along with information on the initial flavour of the decaying hadron, $i.e.$ whether it was produced as a \Bs or a \Bsb meson. The probability density function (PDF) used in the fit consists of signal and background components that include detector resolution and acceptance effects. The PDFs are factorized into separate components for the $\Bsb$ mass and for the remaining observables.

The signal $\BorBbar^0_{\!\!s}$ mass distribution is described by a double Crystal Ball function \cite{Skwarnicki:1986xj}. The background consists of a combinatorial component whose mass distribution is modelled by an exponential function, a 2.3\% contribution from the sum of $\BorBbar^0_{\!\!s}\to \jpsi\eta^\prime$ and $\BorBbar^0_{\!\!s}\rightarrow J/\psi\phi$, with $\phi\to\pi^+\pi^-\pi^0$, and 2.0\% from $B^{\mp}\to \jpsi K^{\mp} + \jpsi  \pi^{\mp}$ decays, both of which produce tails in the $\BorBbar^0_{\!\!s}$ signal region. The latter two background mass shapes are obtained from simulation. The parameters of the signal and the combinatorial background are obtained from a fit to the $\BorBbar^0_{\!\!s}$ mass distribution in an extended region (see Fig.~\ref{fitmass}) and are subsequently fixed for use in the $\phi_s$ fit.

As can be seen from Eqs.~(\ref{Eq-t}) and (\ref{Eqbar-t}), knowledge of the $\BorBbar^0_{\,\,s}$ flavour at production greatly enhances the sensitivity.  The process of determining the initial flavour is called ``tagging". We use both opposite-side~\cite{LHCb-PAPER-2011-027} and same-side tagging information~\cite{Aaij:2013oba}. The opposite-side (OS) tag identifies the flavour of another $b$ hadron in the event using information from the charges of leptons and kaons from its decay, or the charge of another detached vertex. The same-side kaon (SSK) tagger utilizes the hadronization process, where the fragmentation of a $b(\bar{b})$ quark into $\Bsb(\Bs)$ meson can lead to an extra $s(\bar{s})$ quark being available to form a hadron, often leading to a  $K^-(K^+)$ meson. This kaon is correlated to the signal $\BorBbar^0_{\!\!s}$ in phase space, and the sign of its charge identifies the initial flavour.  
 A wrong-tag probability $\eta$ is estimated event-by-event, based on the output of a neural network trained on simulations. 
It is calibrated with data using flavour-specific decay modes in order to predict the true wrong-tag probability of the event $\brabar{\omega}(\eta)$ for an initial flavour $\BorBbar^0_{\!\!s}$ meson, which has a linear dependence on $\eta$.
The calibration is performed separately for the OS and the SSK taggers. 
When events are tagged by both the OS and the SSK algorithms, a combined tag decision and wrong-tag probability are given by the algorithm defined in Ref.~\cite{LHCb-PAPER-2011-027}. This  combined algorithm is implemented in the overall fit. The overall effective tagging power obtained is characterized by $\varepsilon_{\rm tag}D^2=(3.89\pm 0.25)\%$, where $D\equiv(1-2\omega_{\rm avg})$ is the dilution, $\omega_{\rm avg}$ is the average wrong-tag probability, and $\varepsilon_{\rm tag}=(68.68\pm0.33)\%$ is  the signal tagging efficiency.
The overall tagging power is improved by about 60\% with respect to the previous analysis mainly due to the inclusion of the SSK tagger, which has an tagging power about 40\% better than that described in Ref.~\cite{Aaij:2013oba}, due to the use of a neural-network based selection. In addition, the OS algorithms discussed in Ref.~\cite{LHCb-PAPER-2011-027} have been re-optimised using the full available dataset.

The theoretical signal function including flavour tagging is
\begin{align}\label{eq:R}
R(\hat{t},\m,\Omega,\mathfrak{q}|\eta) =&\frac{1}{1+|\mathfrak{q}|}\left[\left[1+\mathfrak{q}\left(1-2\omega(\eta)\right)\right]\Gamma(\hat{t},\m,\Omega)\right.\nonumber\\
&\left. +\left[1-\mathfrak{q}\left(1-2\bar{\omega}(\eta)\right)\right]\bar{\Gamma}(\hat{t},\m,\Omega)\right],
\end{align}
where $\hat{t}$ is the true decay time, and $\brabar{\Gamma}$ is defined in Eqs.~(\ref{Eq-t}) and (\ref{Eqbar-t}). The flavour tag $\mathfrak{q}$ takes values of $-1$, 1, 0, if the signal meson is tagged as $\Bsb$, $\Bs$, or untagged, respectively.

The signal function is further modified to take into account the decay time resolution and the acceptance effects on all the fit variables
\begin{equation}
F^{\rm sig}(t,\m,\Omega,\mathfrak{q}|\eta,\delta_t)= R(\hat{t},\m,\Omega,\mathfrak{q}|\eta) \otimes T(t-\hat{t}|\delta_t) \cdot {\cal E}_t(t) \cdot \varepsilon(\m,\Omega),
\end{equation}
where $\varepsilon(\m,\Omega)$ is the efficiency as a function of $\pi^+\pi^-$ mass and angles, obtained from the simulation as described in Ref.~\cite{Aaij:2014emv}, $T(t-\hat{t}|\delta_t)$ is the decay time resolution function which depends upon the estimated decay time error for each event $\delta_t$, and ${\cal E}_t(t)$ is the decay time acceptance function. The latter two are discussed in Sec.~\ref{sec:time-res-accept}.

The distribution of the background decay time, $\pi^+\pi^-$ mass and angles can be  factorized into components for the decay time and the remaining variables. The background decay time distribution, $F_{t}^{\rm bkg}(t|\delta_t)$ is a double exponential function convolved with the decay time resolution function,  taken to be the same as that of the signal, and multiplied by the background decay time acceptance function. The parameters of the double exponential function and the acceptance function are obtained from the sum of $\jpsi \pi^+\pi^+$ and $\jpsi \pi^-\pi^-$ combinations in the same mass signal window as the $\jpsi \pi^+\pi^-$. The distribution of the background for the $\pi^+\pi^-$ mass and angles is described by the function $B^{\rm bkg}(m_{hh},\Omega)$, discussed in Ref.~\cite{Aaij:2014emv}, by summing all the backgrounds components.

The events are divided into four tagging categories: tagged by both OS and SSK, by OS only, by SSK only, and untagged. Each category ${i}$ is described by the PDF
\begin{align}
P^{i}&(m,t, m_{hh},\Omega,\eta,\mathfrak{q}, \delta_{t})  \nonumber\\
&=\frac{ (1-f^{i}_{\rm bkg})} {{\cal N}^{i}_{\rm sig} } P_m^{\rm sig}(m)F^{\rm sig}(t,m_{hh},\Omega,\mathfrak{q}|\eta,\delta_{t})  P_{\delta_t}^{\rm sig}(\delta_{t})  P_{\eta,i}^{\rm sig}(\eta) \nonumber\\
&+\frac{ f^{i}_{\rm bkg}}{{\cal N}^{i}_{\rm bkg}}  P_m^{\rm bkg}(m)  B^{\rm bkg}(m_{hh},\Omega) F^{\rm bkg}_{ t}(t|\delta_{t})  P_{\delta_t}^{\rm bkg}(\delta_{t}) P_{\eta,i}^{\rm bkg}(\eta),
\end{align}
where $f^{i}_{\rm bkg}$ is the background fraction, which is fixed to the  value obtained from the $\BorBbar^0_{\!\!s}$ mass fit for each category. The normalization factors ${\cal N}^{i}$ are calculated for each tag category by integrating over the decay time $t$, the dihadron invariant mass $\m$, and the angles $\Omega$. 

\begin{figure}[tb]
\begin{center}
\includegraphics[width=0.5\textwidth]{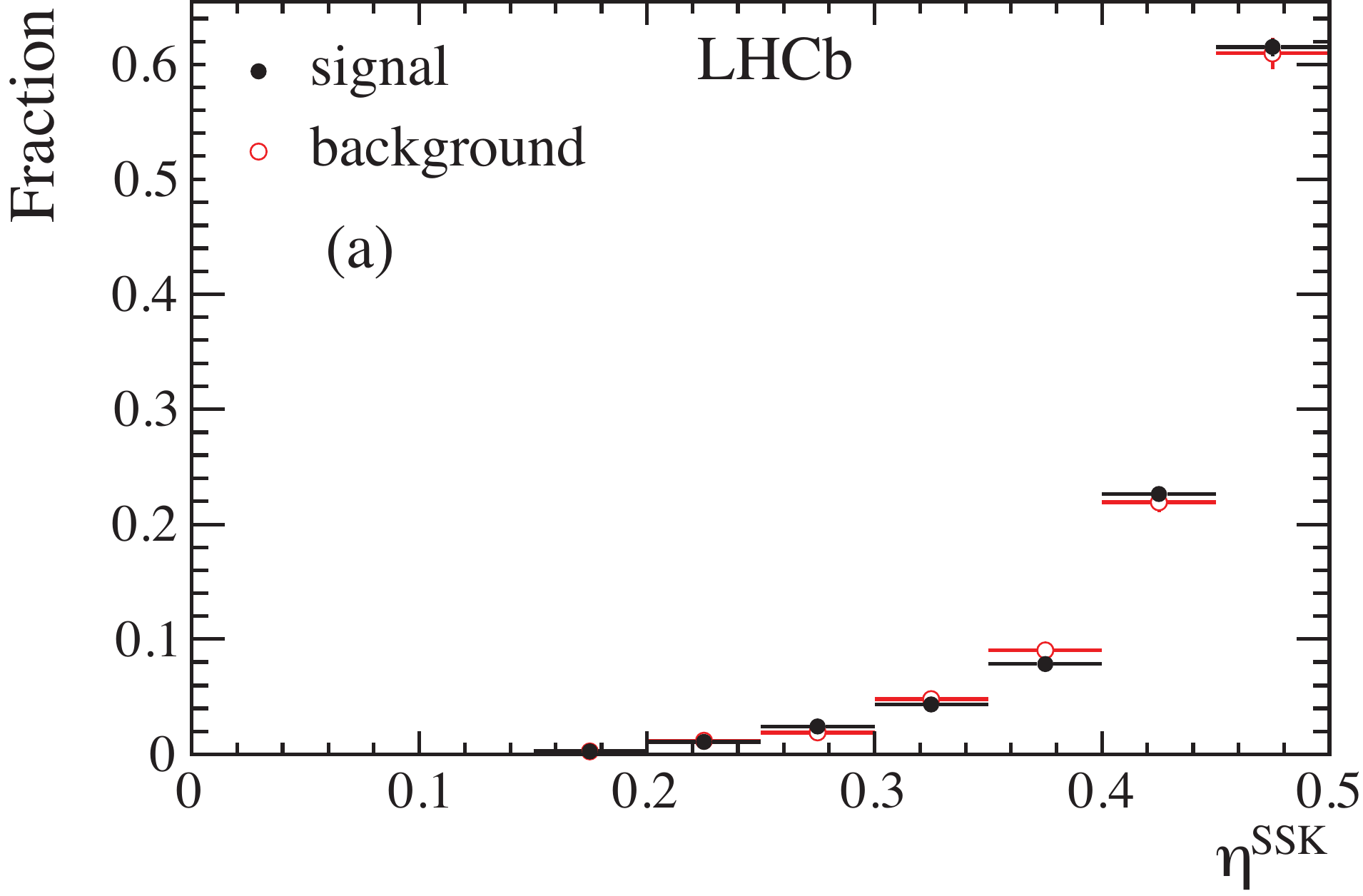}%
\includegraphics[width=0.496\textwidth]{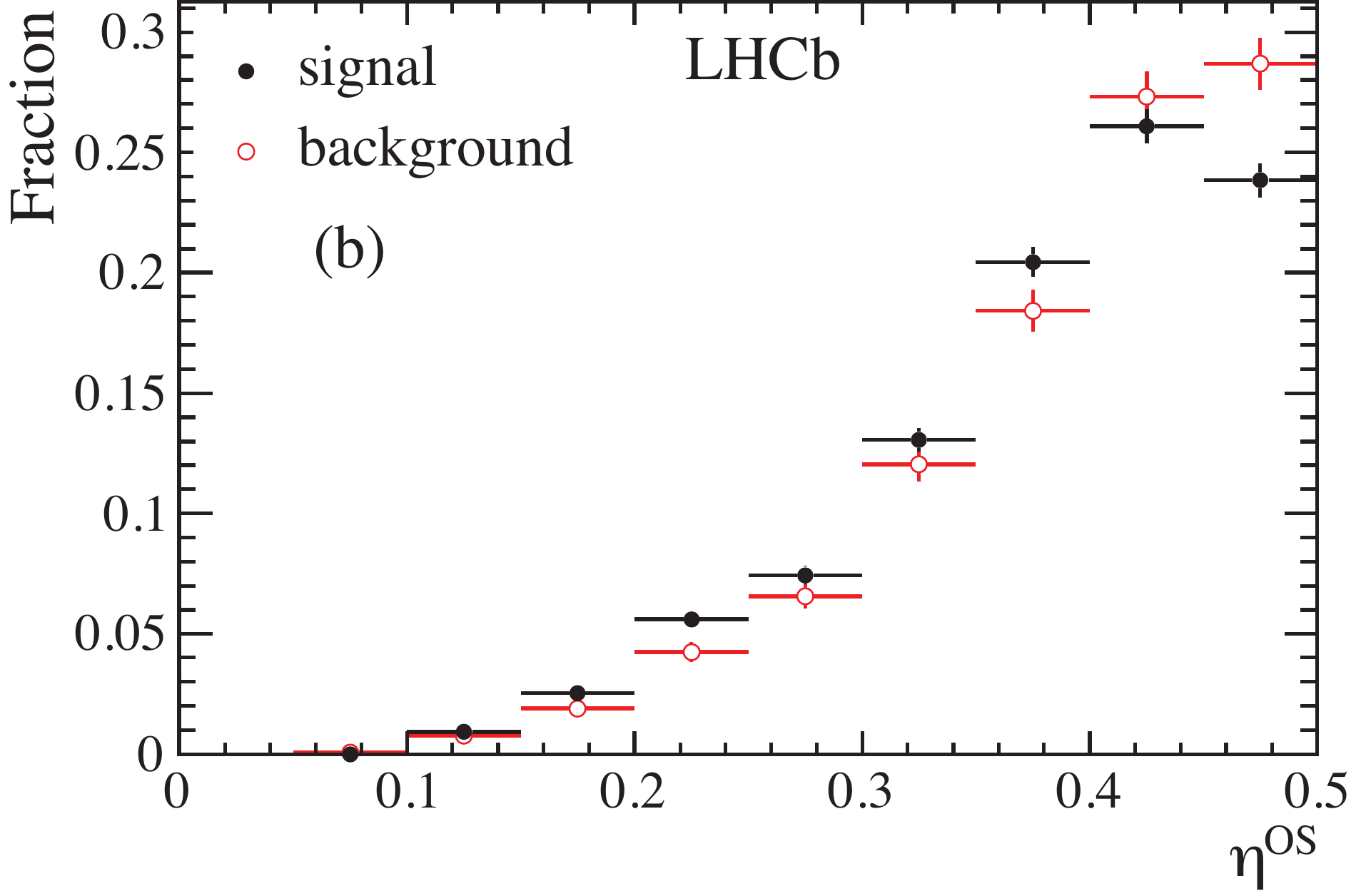}
\end{center}\label{eta1d}
\vspace{-4mm}
\caption{\small Estimated fractions of mistag probabilities from (a) the SSK tagger, $\eta^{\rm SSK}$, and (b) the OS tagger, $\eta^{\rm OS}$. }
\end{figure}

We include the PDFs for the estimated per-candidate decay time error $\delta_t$ and the wrong-tag probability $\eta$. The $P_{\delta_t}^{\rm sig}(\delta_t)$ and $P_{\delta_t}^{\rm bkg}(\delta_t)$ functions are signal and background PDFs of $\delta_t$. The background PDF is obtained from the distribution of the like-sign events and the signal PDF is obtained from the distribution of the $\BorBbar^0_{\!\!s}$ candidates after background subtraction. The signal peaks at about 26\fs and the background at 29\fs. The mistagging PDF is different in each of the tagging categories: it is a product of two one-dimensional PDFs of $\eta^{\rm SSK}$ and $\eta^{\rm OS}$ if both are tagged, a one-dimensional PDF of the corresponding tagger if only single tagged, and a uniform PDF  if untagged. The two one-dimensional distributions of $\eta^{\rm SSK}$ and $\eta^{\rm OS}$ are shown in Fig.~\ref{eta1d} for both signal and background.

\section{Decay time resolution and acceptance}
\label{sec:time-res-accept}
The decay time resolution function $T(t-\hat{t};\delta t)$ is described by a sum of three Gaussian functions with a common mean,  and widths given by three scale factors, each being multiplied by $\sigma_t \equiv \delta_t + \sigma_t^0$, where $\delta_t$ is the estimated per-event decay time error and $\sigma_t^0$ is a constant parameter. Studies on simulated data show that prompt $\jpsi \pi^+\pi^-$ combinations have nearly identical resolution to signal events. Consequently, we determine the parameters of the resolution model from a fit to the decay time distribution of such prompt combinations in the data, where the contribution of candidates unlikely to originate from $\jpsi$ events are subtracted.
Taking into account the $\delta_t$ distribution of the $\BorBbar^0_{\!\!s}$ signal, the effective resolution is found to be 40.3 fs.

The decay time distribution is influenced by acceptance effects that are introduced by track reconstruction,  trigger and  event selection. The decay time acceptance is obtained using control samples of $\Bzb \to J/\psi \Kstarzb(\to K^-\pi^+)$ and $\Bz \to J/\psi \Kstarz(\to K^+\pi^-)$  decays, and then corrected by the acceptance ratio between $\Bsb$ and $\Bzb$ decays derived from the simulation.

\begin{figure}[!b]
\begin{center}
\includegraphics[width=0.461\textwidth]{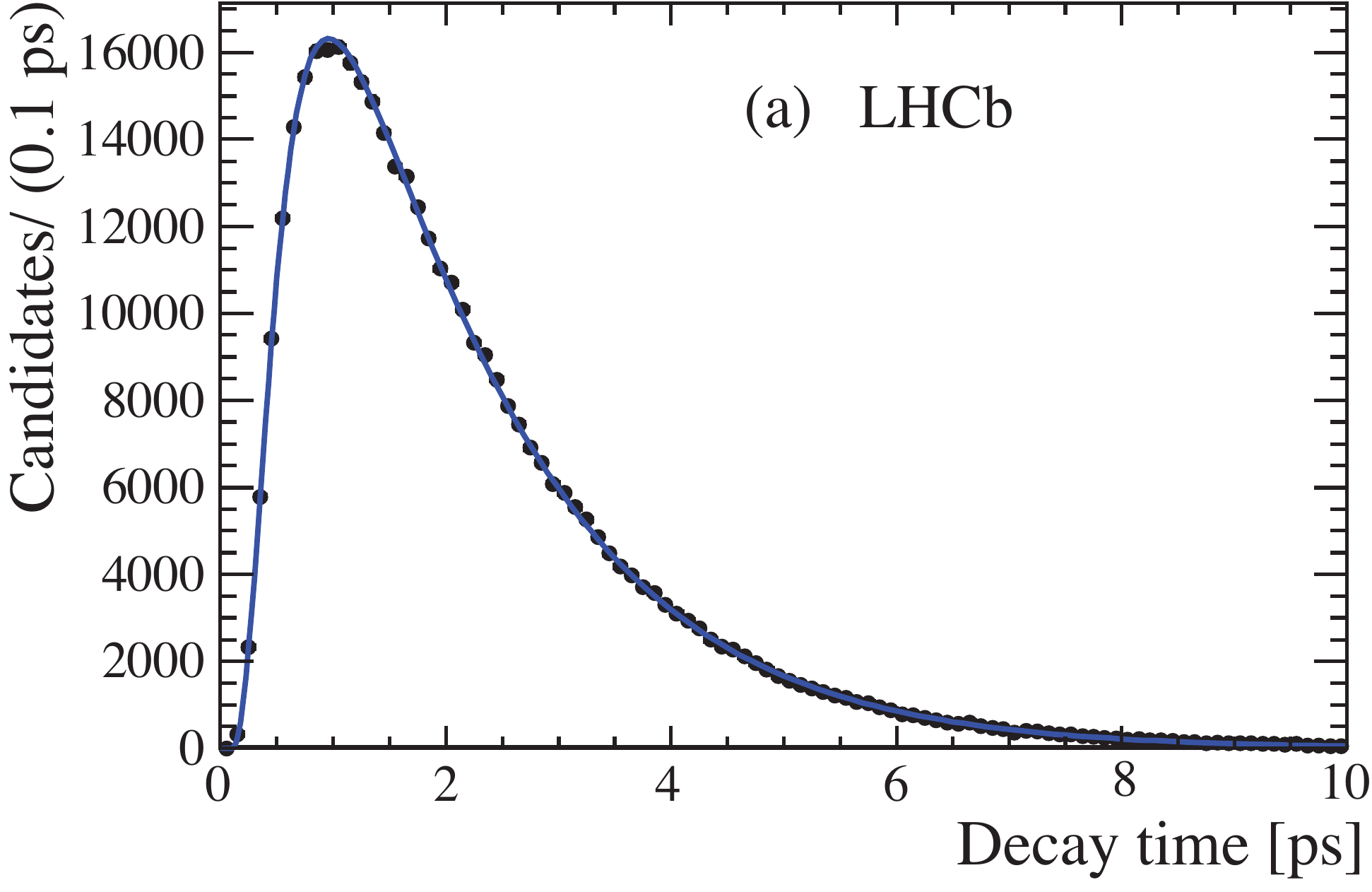}
\includegraphics[width=0.50\textwidth]{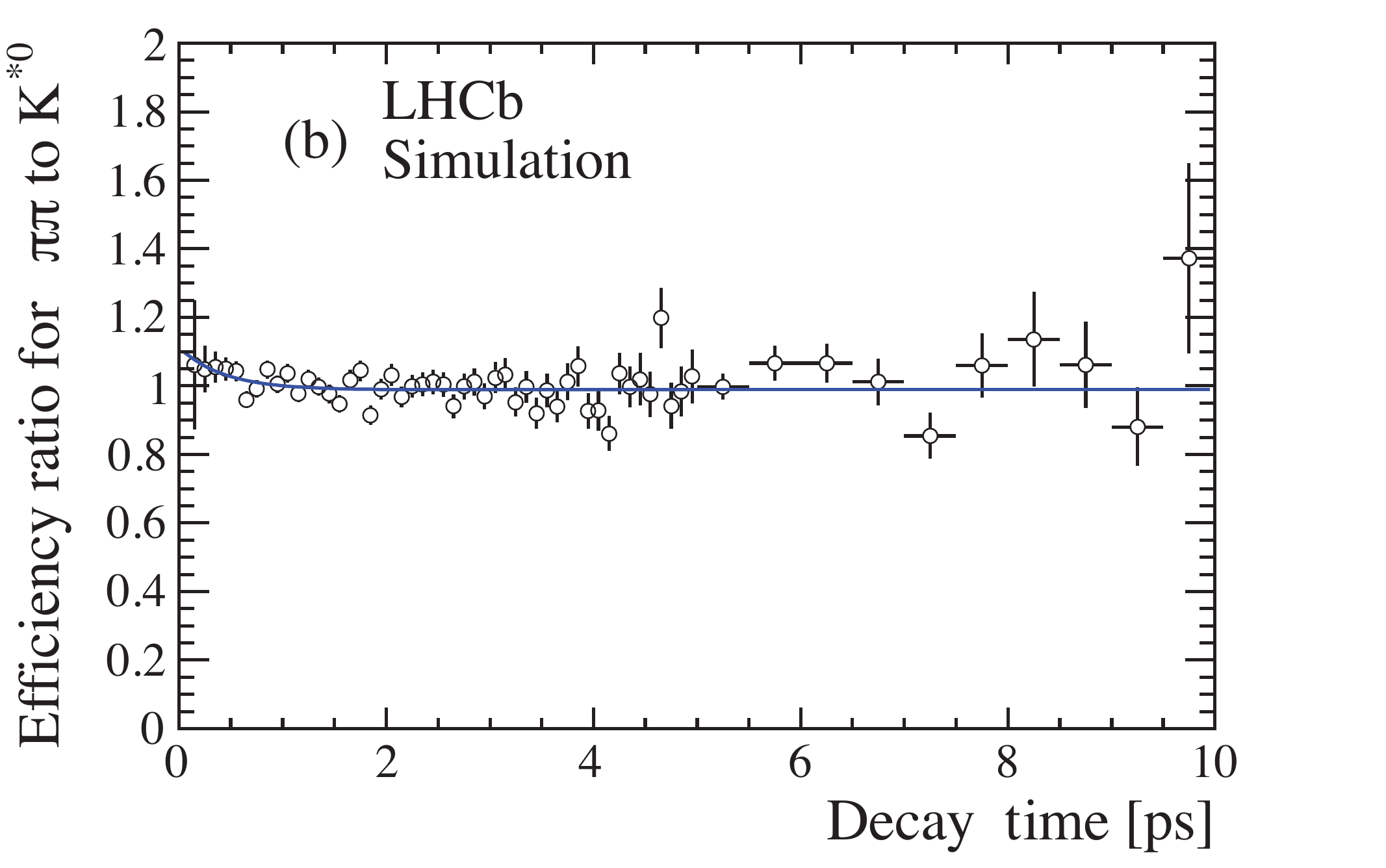}
\end{center}\label{fitkst}
\vspace{-4mm}
\caption{\small Distributions of (a) decay time of $\BorBbar^0 \to J/\psi \KorKbar^{\!*0}$ candidates in data, (b) ratio of time acceptance between $\Bsb\to \jpsi \pi^+\pi^-$ and $\Bzb \to J/\psi \Kstarzb$ decays from simulation.}
\end{figure}

The same selection  as for signal events is implemented for the $\BorBbar^0$ candidates except for the kaon identification requirement. The $K^{\mp}\pi^{\pm}$ pair mass is restricted within $\pm100$\,MeV of the nominal $\Kstarzb$ mass~\cite{PDG}. The candidates within $\pm25$\mev of the $\Bzb$ mass peak are used to measure the decay time acceptance. There are $399\,200 \pm 800$ signal events with a purity of 98.5\%. The decay time distribution is shown in Fig.~\ref{fitkst}(a). These data are fitted with an exponential function convolved with the time resolution function, and then multiplied by the acceptance function, $\frac{\left[a(t-t_0)\right]^n}{1+\left[a(t-t_0)\right]^n} \times (1+\beta t+\beta_2 t^2)$,
where $a$, $n$, $t_0$, $\beta$, and $\beta_2$ are parameters determined by the fit. The $\Bzb$ lifetime is constrained to $\tau_{\Bz}=1.519\pm0.007$~ps~\cite{PDG}.
The signal acceptance parameters and their correlations are given in Table \ref{acc-par}. There is a large efficiency drop below 1\ps due to detachment requirements on the $\BorBbar^0$ and its decay products in the selection.

Figure~\ref{fitkst}(b) shows the acceptance ratio between $\Bsb\to J/\psi \pi^+\pi^-$ and $\Bzb\to J/\psi \Kstarzb$ decays from the simulation. The distribution is almost flat. The ratio is well described by the function $R(1-p_2 e^{-p_1 t})$ with parameters $R$, $p_1$ and $p_2$ determined by the fit. Parameter $R$ is a normalization constant.

We use the product of the acceptance determined from $\BorBbar^0 \to J/\psi \KorKbar^{\!*0}$ decays  and the correction ratio found from simulation as the decay time acceptance function for $\Bsb$, denoted as
${\cal E}_t(t;a,n,t_0,\beta,\beta_2,p_1,p_2)$, 
where the parameter values and correlations are given in Table~\ref{acc-par}. 
\begin{table}
\caption{\small Acceptance function parameter values and their correlations.}
\centering
\begin{tabular}{lrrrrrccc}\hline
\multicolumn{8}{c}{Parameter correlations} &  Values\\
 &$n~~$ & $a~~$ & $\beta$~~ & $\beta_2$~ & $t_0$~~ & $p_1$ & $p_2$ &\\\hline
$n$ &~~1.00 & ~~0.44 & ~~0.57& $-$0.54 &$-$0.86& ~~0.00 & ~~0.00 & ~~$2.082\pm0.036$\,~~~~~~~\\
 $a$ &~~0.44 &~~ 1.00 &~~ 0.74 &$-$0.74 &$-$0.05 & ~~0.00 & ~~0.00 & ~~$1.981\pm0.024$\,ps$^{-1}$\\
$\beta$  &~~0.57 & ~~0.74 & ~~ 1.00 &$-$0.90 &$-$0.37& ~~0.00 & ~~0.00 & ~~$0.077\pm0.009$\,ps$^{-1}$\\
$\beta_2$ &$-$0.54 &$-$0.74 &$-$0.90 & ~~1.00&  ~~0.34& ~~0.00 & ~~0.00 & $-0.008\pm0.001$\,ps$^{-2}$\\
 $t_0$ &$-$0.86&$-$0.05 &$-$0.37 & ~~0.34&  ~~1.00& ~~0.00 & ~~0.00 & ~~$0.104\pm0.003$\,ps~~~~\\
  $p_1$ &~~ 0.00 & ~~0.00 &~~ 0.00 & ~~0.00 & ~~0.00 & ~~1.00 & $-$0.89 & ~~$2.290\pm1.761$\,ps$^{-1}$\\
$p_2$ &~~ 0.00 & ~~0.00 & ~~0.00 & ~~0.00 & ~~0.00 &  $-$0.89 & ~~1.00 & $-0.124\pm0.110$\,~~~~~~~\\\hline
\end{tabular}\label{acc-par}
\end{table}

\section{Results}
The \CP phase $\phi_s$ is determined from the fit that uses the amplitude model with five final state $\pi^+\pi^-$ resonances.
Several of the model parameters have  Gaussian constraints applied in the fit. They are the  measured values of $\dms=17.768\pm0.024$ ps$^{-1}$~\cite{Aaij:2013mpa}, $\Gs=0.663\pm0.005\pm0.006$ ps$^{-1}$ and $\DGs=0.100\pm0.016\pm0.003$ ps$^{-1}$~\cite{Aaij:2013oba}, the tagging parameters, 
the mass and width of the $f_0(1790)$ \cite{Ablikim:2004wn},  the $f_2^\prime(1525)$ fit fractions, and the scale factors in the decay time resolution function, multiplied by  $(1.00\pm0.05)$ to take into account the systematic uncertainty on the decay time resolution estimate~\cite{LHCb:2012ad}. Apart from $\phi_s$ and $|\lambda|$, the other free parameters are the amplitudes and phases of the $\pi^+\pi^-$ states. The fit procedure is checked using pseudoexperiments with the same size as data. The fit reproduces the input $\phi_s$ values with negligible bias. 

\begin{figure}[t]
\begin{center}
\includegraphics[width=0.65\textwidth]{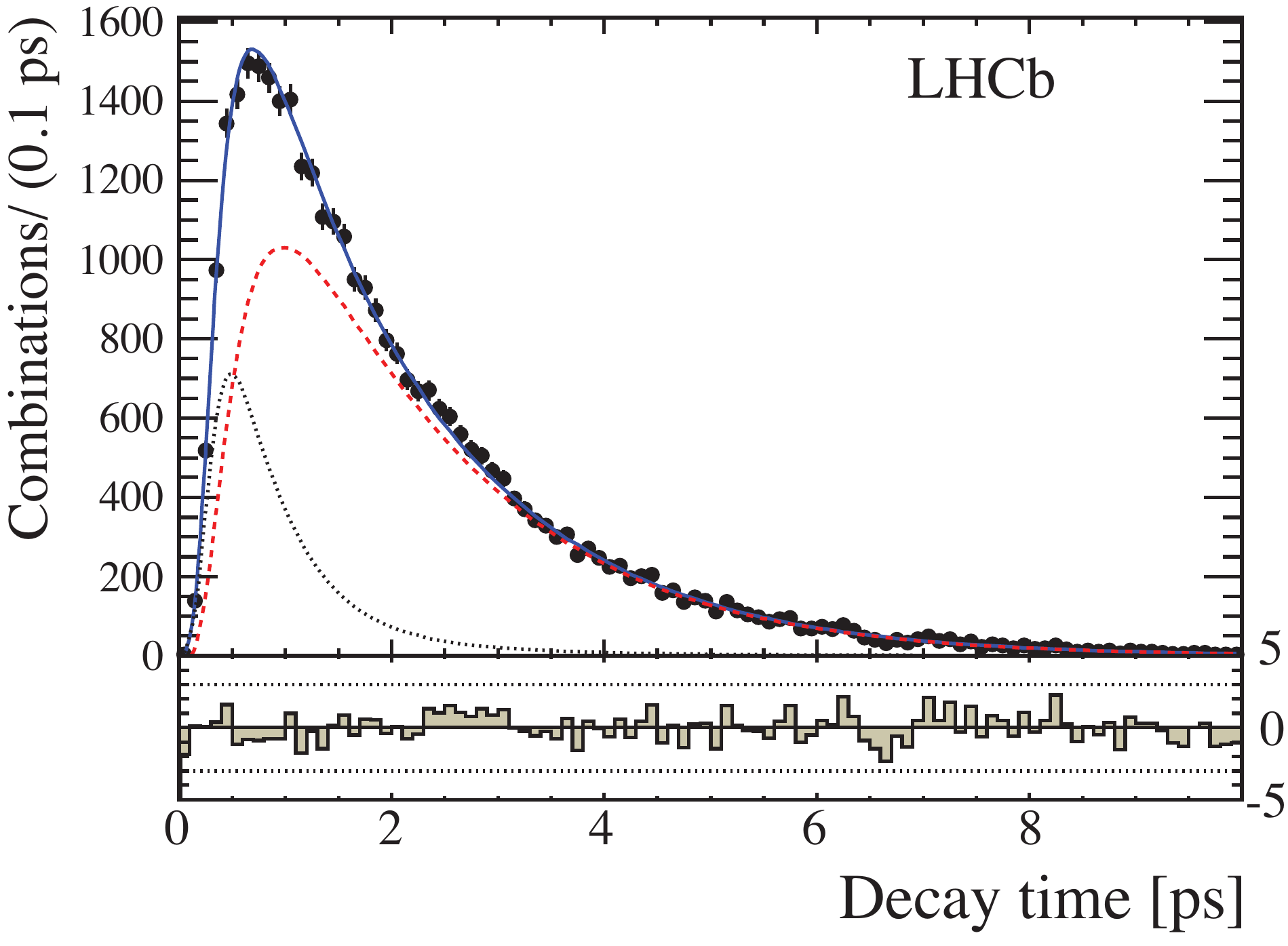}
\caption{\small Decay time distribution of $\BorBbar^0\to\jpsi \pi^+\pi^-$ candidates. The signal PDF is shown with a (red) dashed line, the background with a (black) dotted line, and the (blue) solid line represents the total. }\label{finalfit}
\end{center}
\end{figure}

For our first fit we do not allow  direct \CP violation and therefore fix $|\lambda|$ to 1. The fit determines $\phi_s=75\pm 67\pm 8$\,mrad. When two uncertainties are quoted, the first is statistical and the second the systematic. The systematic uncertainty is discussed in Sec.~\ref{sec:sys}.  Figure~\ref{finalfit} shows the decay time distribution superimposed with the fit projection.  Projections for $\m$ and $\Omega$ are shown in Fig.~\ref{RM8}.
Fit fractions of the contributing resonances are consistent with the results from the amplitude analysis~\cite{Aaij:2014emv}. We also perform the fit with $|\lambda|$ treated as a free parameter. The fit determines $\phi_s=70\pm68\pm8$\,mrad and $|\lambda|=0.89\pm0.05\pm0.01$, consistent with no direct \CP violation ($|\lambda|=1$), under the assumption that direct \CP violation is equal for all of the intermediate $\pi^+\pi^-$ states. (The correlation between $\phi_s$ and $|\lambda|$ is about 1\%.)

Since the $\jpsi \pi^+\pi^-$ final state is known to be $>$97.7\% CP-odd at 95\% CL~\cite{Aaij:2014emv}, we check our result by implementing a simplified fit without using the information of $m_{hh}$ and $\Omega$. Here the \CP-odd fraction is assumed to be 100\%, thus angular information is not needed to separate \CP-odd and possible \CP-even components. This fit was used in the previous $\phi_s$ measurement using $\jpsi \pi^+\pi^-$ decays~\cite{LHCb:2012ad}. Compared to the fit discussed above, the simplified fit gives a $\phi_s$ value differing by $20$\,mrad and a statistical uncertainty of $\pm$69\,mrad. The small difference between the two fits is consistent with a study using  pseudoexperiments, where the distribution of the difference between the two fits is a Gaussian with a mean of zero and a width of 20\,mrad. 

\section{Systematic uncertainties}
\label{sec:sys}
The systematic uncertainties on $\phi_s$ and $|\lambda|$, evaluated using the fit allowing direct \CP-violation, are summarized in Table~\ref{tab:sys}. They are small compared to the statistical uncertainty. Since Gaussian constraints are applied in the fit, no additional uncertainty is introduced by the input parameters $\dms$, $\Gs$, $\DGs$, or those associated with flavour tagging and time resolution. 

To evaluate the uncertainties due to the fixed parameters in the decay time acceptance, background decay time PDF, $m(\pi^+\pi^-)$ and $m(\jpsi\pi^{\pm})$ (mass) acceptance and background mass PDF, the data fit is repeated by varying the fixed parameters from their nominal values according to the error matrix 200 times for each source. The matrix elements are determined using simulation, $\jpsi \overline{K}^*$ data, and like-sign dipion data. The r.m.s.\,of the fitted $\phi_s$ value is taken as the uncertainty for each source. 

Including different resonances could change the \CP-even fraction in the decay, and thus the $\phi_s$ result. In Ref.~\cite{Aaij:2014emv} two acceptable solutions were found for the contributing components. For our main result we use the one with five resonant components. The other solution adds a 5.9\% nonresonant component. Evaluating  $\phi_s$ for the second solution gives a small difference of $3$\,mrad. Adding a $\rho(770)$ component causes the largest change for $\phi_s$ and $\lambda$ and is taken as the systematic uncertainty, even though vector particles must conserve the zero isospin of the dipion system, which forbids the decay into $\rho(770)$. The resonance masses and widths of $f_2(1270)$ and $f_2^\prime(1525)$ are fixed in the fit. 

To evaluate the uncertainty due to the fixed masses and widths, the fit is repeated by changing each parameter within one standard deviation of its error, and the larger shift in the fitted values is taken as the systematic uncertainty. Similarly, the uncertainties due to other fixed parameters, such as background fractions and those used in $\BorBbar$ mass PDFs, are also determined.
We take the background decay time distribution to be independent of $\m$. This assumption is tested by repeating the fit with different background decay time PDFs for the low $\m$ and high $\m$ regions, found from the like-sign dipion events in the same mass regions. The effects on $\phi_s$ and  $|\lambda|$ are found to be negligible.

The production ratio of $\Bsb$ to $\Bs$ is estimated to be  $R_{\rm P}=(1.00\pm0.05)$ \cite{Norrbin:2000jy}. To include this effect, the $\Bsb$ decay rate, $\bar{\Gamma}$, used in Eq.~(\ref{eq:R})  is multiplied by $R_{\rm P}$. The uncertainty due to this source is estimated by varying $R_{\rm P}$ within its error. The uncertainties are added in quadrature to give the total.


\begin{table}[bt]
\begin{center}
\caption{\small Systematic uncertainties. The total is the sum in quadrature of each entry.\label{tab:sys}}
\begin{tabular}{lcc}\hline
Sources &  $\phi_s$(mrad) & $\lambda$\\\hline
Decay time acceptance  & $\pm0.6$ & $\pm0.0008$\\
Mass acceptance & $\pm0.3$ & $\pm0.0003$\\
Background time PDF&  $\pm0.2$ & $\pm0.0011$\\
Background mass distribution PDF& $\pm0.6$ & $\pm0.0016$\\
Resonance model &  $\pm6.0$ & $\pm0.0100$\\
Resonance parameters& $\pm0.7$ &$\pm0.0007$\\
Other fixed parameters& $\pm0.4$ &$\pm0.0009$\\
Production asymmetry& $\pm5.8$& $\pm0.0017$\\\hline
Total &  $\pm8.4$ & $\pm0.010$\\\hline
\end{tabular}
\end{center}
\end{table}

\section{Conclusions}
We have presented a  time-dependent flavour-tagged analysis of the $\BorBbar^0_{\!\!s}\to\jpsi\pi^+\pi^-$ decay using  angular distributions and  the $\pi^+\pi^-$ mass dependence to determine the \CP content of the final state components. We measure the mixing induced \CP-violating phase $\phi_s$. Assuming the absence of direct \CP violation, we find
\begin{equation}
\phi_s=75\pm 67\pm8~{\rm mrad}.\nonumber
\end{equation}
For the case where direct \CP is allowed, we find
\begin{equation}
\phi_s=70\pm 68\pm 8~{\rm mrad},~|\lambda|=0.89\pm 0.05\pm 0.01.\nonumber
\end{equation}
This result supersedes and is more precise than our previous measurement in this decay mode of
$\phi_s=-19^{+173+4}_{-174-3}$\,mrad based on a 1\,\invfb data sample~\cite{LHCb:2012ad}.
Physics beyond the Standard Model is not established by our measurements.

\section*{Acknowledgements}
 
\noindent We express our gratitude to our colleagues in the CERN
accelerator departments for the excellent performance of the LHC. We
thank the technical and administrative staff at the LHCb
institutes. We acknowledge support from CERN and from the national
agencies: CAPES, CNPq, FAPERJ and FINEP (Brazil); NSFC (China);
CNRS/IN2P3 and Region Auvergne (France); BMBF, DFG, HGF and MPG
(Germany); SFI (Ireland); INFN (Italy); FOM and NWO (The Netherlands);
SCSR (Poland); MEN/IFA (Romania); MinES, Rosatom, RFBR and NRC
``Kurchatov Institute'' (Russia); MinECo, XuntaGal and GENCAT (Spain);
SNSF and SER (Switzerland); NASU (Ukraine); STFC and the Royal Society (United
Kingdom); NSF (USA). We also acknowledge the support received from EPLANET,
Marie Curie Actions and the ERC under FP7.
The Tier1 computing centres are supported by IN2P3 (France), KIT and BMBF (Germany),
INFN (Italy), NWO and SURF (The Netherlands), PIC (Spain), GridPP (United Kingdom).
We are indebted to the communities behind the multiple open source software packages on which we depend.
We are also thankful for the computing resources and the access to software R\&D tools provided by Yandex LLC (Russia).

\input{phis-Bs2Jpsipipi-paper.bbl}

\end{document}

%% file: LHCb_HD_authorlist_2014-04-08.tex
\centerline{\large\bf LHCb collaboration}
\begin{flushleft}
\small
R.~Aaij$^{41}$, 
B.~Adeva$^{37}$, 
M.~Adinolfi$^{46}$, 
A.~Affolder$^{52}$, 
Z.~Ajaltouni$^{5}$, 
S.~Akar$^{6}$, 
J.~Albrecht$^{9}$, 
F.~Alessio$^{38}$, 
M.~Alexander$^{51}$, 
S.~Ali$^{41}$
% ...

{\footnotesize \it
% ...
$ ^{5}$Clermont Universit\'{e}, Universit\'{e} Blaise Pascal, CNRS/IN2P3, LPC, Clermont-Ferrand, France\\
$ ^{6}$CPPM, Aix-Marseille Universit\'{e}, CNRS/IN2P3, Marseille, France\\
% ...
$ ^{9}$Fakult\"{a}t Physik, Technische Universit\"{a}t Dortmund, Dortmund, Germany\\
% ...
$ ^{37}$Universidad de Santiago de Compostela, Santiago de Compostela, Spain\\
$ ^{38}$European Organization for Nuclear Research (CERN), Geneva, Switzerland\\
% ...
$ ^{41}$Nikhef National Institute for Subatomic Physics, Amsterdam, The Netherlands\\
% ...
$ ^{46}$H.H. Wills Physics Laboratory, University of Bristol, Bristol, United Kingdom\\
% ...
$ ^{51}$School of Physics and Astronomy, University of Glasgow, Glasgow, United Kingdom\\
$ ^{52}$Oliver Lodge Laboratory, University of Liverpool, Liverpool, United Kingdom\\
% ...
}
\end{flushleft}

%% file: phis-Bs2Jpsipipi-paper.bbl
\ifx\mcitethebibliography\mciteundefinedmacro
\PackageError{LHCb.bst}{mciteplus.sty has not been loaded}
{This bibstyle requires the use of the mciteplus package.}\fi
\providecommand{\href}[2]{#2}